\documentclass[twocolumn,prl]{revtex4}%
\usepackage{graphicx}%
\usepackage{amsmath}%
\usepackage{amsfonts}%
\usepackage{amssymb}
\usepackage{bm}

\def\d{{\partial}}
\def\s{{\sigma}}

\def\k{{ {\bm k} }}

\begin{document}
\title{Comment on ``Isoelectronic Ru substitution at Fe-site in 
Sm(Fe$_{1-x}$Ru$_x$)AsO$_{0.85}$F$_{0.15}$ compound and its effects 
on structural, superconducting and normal state properties''
(arXiv:1004.1978)
}
\author{Hiroshi \textsc{Kontani}$^{1}$, and Masatoshi \textsc{Sato}$^{1,2}$}

\date{\today }

\begin{abstract}
Based on the five-orbital model, 
we derive the reduced impurity scattering rate 
$g=z\gamma/2\pi T_{\rm c0}$ in Sm(Fe$_{1-x}$Ru$_{x}$)AsO$_{0.85}$F$_{0.15}$ 
from the residual resistivity.
At $x=0$, the transition temperature is $T_{\rm c0}=50$ K.
For $0.05\le x \le 0.36$ ($0.84 \ge T_{\rm c}/T_{\rm c0} \ge 0.3$) 
the obtained value of $g$ ranges from $1.5$ to $2.9$,
which suggests that the $s_\pm$-wave state cannot survive.
We point out that the magnetoresistance frequently
gives an underestimated value of $g$ in correlated electron systems.
\end{abstract}

\address{
$^1$ Department of Physics, Nagoya University and JST, TRIP, 
Furo-cho, Nagoya 464-8602, Japan. 
\\
$^2$ Toyota Physical and Chemical Institute and JST, TRIP,
Nagakute, Aichi 480-1192, Japan
}
 

\sloppy

\maketitle

In Ref. \cite{Tropeano}, it was shown that 
the transition temperature in polycrystal 
Sm(Fe$_{1-x}$Ru$_x$)AsO$_{0.85}$F$_{0.15}$ at $x=0$, $T_{\rm c0}=50$ K,
decreases to $T_{\rm c}=15$ K at $x=0.36$,
and $T_{\rm c}=0$ K at $x=0.75$.
The observed weak $x$-dependence of $T_{\rm c}$
is consistent with previous reports \cite{Sato,alpha}.
The authors discussed the reduced scattering rate $g=z\gamma/2\pi T_{\rm c0}$ 
where $\gamma$ is the electron scattering rate
and $z=m/m^*$ is the renormalization factor.
Theoretically, the sign reversal $s$-wave ($s_\pm$-wave) state 
vanishes for $g>g_c^{s\pm}=0.23$
 \cite{Onari-impurity};
$g_c^{s\pm}\ll1$ due to large interband impurity scattering,
since all bands are composed of the same $d$-orbitals.
Thus, it is essential to derive $g$ from experiments.

Here, we derive a reliable value of $g$ 
from the the residual resistivity $\rho_{\rm 0}$,
since we can calculate $\rho_{\rm 0}$ 
with enough accuracy due to the fact that
the realistic five-orbital model is available \cite{Kuroki}.
In the linear-response theory, 
$\s_{xx}=\frac{e^2}{4\pi c}\sum_i \int_{{\rm FS}i}dS_\k
 \frac{|{\bm v}_\k^i|}{2\gamma}$ \cite{Onari-impurity},
where $i$ represents the Fermi surfaces, ${\bm v}_\k^i=\nabla_\k E_\k^i$,
and $c$ is the inter-layer spacing.
Using the five-orbital model, we obtain 
the relation $\rho {\rm [\mu\Omega cm]} = 0.24 \gamma {\rm [K]}$
in 1111 systems ($c=0.8$nm) for the filling $n=5.8-6.1$.
Then, $g$ is given as
$g= 0.66z\cdot\rho_{\rm 0} [\mu\Omega{\rm cm}]/T_{\rm c0}[{\rm K}]$.
In the derivation, correct physical quantities,
such as carrier density and Fermi velocity,
are maintained automatically.
In the {\it single crystal} NdFeAsO$_{0.7}$F$_{0.3}$ with $T_{\rm c0}=46.4$ K,
$T_{\rm c}$ decreases to $T_{\rm c0}/2$
by $\alpha$-particle irradiation when 
$\rho_{\rm 0}\sim 480 \ \mu\Omega{\rm cm}$.
Thus, $T_{\rm c}$ is halved when $g=3.4=15g_c^{s\pm}$ for $z=1/2$.

Next, we discuss Sm(Fe$_{1-x}$Ru$_x$)AsO$_{0.85}$F$_{0.15}$.
According to Fig. 15 of Ref. \cite{Tropeano}, 
$\rho_{\rm 0}^{\rm poly}\sim 900\ \mu\Omega{\rm cm}$ at $x=0.05$ ($T_{\rm c}=42$K),
which corresponds to $\rho_{\rm 0}^{\rm single}\sim 220\ \mu\Omega{\rm cm}$
in the single crystal, if we apply the empirical relation 
$\rho^{\rm poly}/\rho^{\rm single}\sim4$ \cite{Sato}.
Since the corresponding $g$ is 1.5 for $z=1/2$,
$s_\pm$-wave state cannot survive against 5\% Ru impurities,
unless $\rho^{\rm poly}/\rho^{\rm single}>26$.
These results are consistent with the previous report for Nd1111 \cite{Sato}.
We also obtain $g\sim 2.9$ at $x=0.36$ ($T_{\rm c}=15$ K)
where $\rho_{\rm 0}^{\rm poly}\sim 1750\ \mu\Omega{\rm cm}$.
However, the reduced scattering rates derived from
the magnetoresistance $\Delta\rho/\rho_0\equiv (\mu_{\rm MR}B)^2$ 
and Hall angle $\s_{xy}/\s_{xx}\equiv\mu_{\rm H}B$ at $x=0.36$ 
in Ref. \cite{Tropeano}
are $g_{\rm MR}\sim0.5$ and $g_{\rm H}\sim7.5$ at 57K, respectively \cite{comment}.
For $0.05\le x\le0.33$, $g_{\rm MR}=0.9-1.5$ and 
$g_{\rm H}=1.1-7.5$ according to Fig. 12 of Ref. \cite{Tropeano}.
Thus, both $g_{\rm MR}$ and $g_{\rm H}$ exceed $g_c^{s\pm}$ for $x\ge0.05$.

It should be noticed that $\mu_{\rm H, MR}$ 
can prominently deviate from the true mobility in correlated metals.
For example, the ratio $r\equiv\mu_{\rm MR}/\mu_{\rm H}$ is not unity in general.
The inset of Fig. 3(b) in Ref. \cite{Matsuda} shows that the relation 
$\Delta\rho/\rho_0/(\s_{xy}/\s_{xx})^2=r^2\approx 9$ holds in single crystal 
BaFe$_2$(As,P)$_2$, suggesting that $r\approx 3$.
In high-$T_{\rm c}$ cuprates, $r^2\sim3$ in Y- and Bi-based compounds,
whereas $r^2\approx 14$ in La$_{1-x}$Sr$_{x}$CuO$_4$ at $x=0.17$ \cite{Ong},
and $r^2\sim 100$ at $x\gtrsim0.23$ \cite{LSCO2}.
Thus, the material dependence of $r=\mu_{\rm MR}/\mu_{\rm H}$ 
is very large even in systems with single Fermi surface \cite{ROP}.
[For example, $r$ becomes large when mean-free-path is anisotropic
 \cite{comment2}.]
Moreover, $\mu_{\rm H,MR}$ can deviate from the true mobility 
below $\sim200$ K due to the current vertex correction (CVC)
in the presence of strong spin (or orbital) fluctuations \cite{ROP}.
In fact, Ref. \cite{Matsuda} revealed the significant
role of the CVC in BaFe$_2$(As,P)$_2$.

To summarize, we derived the expression
$g= 0.66z\cdot\rho_{\rm 0}^{\rm single}[\mu\Omega{\rm cm}]/T_{\rm c0}[{\rm K}]$,
which is reliable since $\rho_{\rm 0}/\gamma$ essentially depends only on 
the averaged $k_{\rm F}$ and $v_{\rm F}$ for each Fermi surface,
whereas insensitive to the correlation effects.
For the $s_\pm$-wave state, the critical residual resistivity for 
$g_c^{s\pm}=0.23$ is only $35\ \mu\Omega{\rm cm}$ if $z\sim1/2$. 
This fact supports the $s_{++}$-wave state
predicted by the orbital-fluctuation theory \cite{Kontani}.
Slow decrease in $T_{\rm c}$ ($-\Delta T_{\rm c}\sim 2 K$ per 1\% impurities) 
might originate from the localization effect or 
reduction of orbital fluctuations by impurities.
In contrast to $\rho_0$, Hall angle and $\Delta\rho/\rho_0$ 
are not simply scaled by $\gamma$ since they are sensitive to 
other factors, like the Fermi velocity anisotropy and the CVC.
Therefore, we have to take care in deriving $g$ from them.



\begin{thebibliography}{99}
\bibitem{comment2}
In the linear response theory,
$\s_{xy}$ and the magneto-conductivity are respectively given by
$\s_{xy}\sim B\sum_i\langle|{\bm l}_\k|^2(\d\theta_\k/\d k_\parallel)\rangle_{{\rm FS}i}$
and $\Delta\s_{xx} \sim -B^2\sum_i
\langle ({\bm l}_\k^2(\d\theta_\k/\d k_\parallel)^2 \gamma_\k^{-1}
+(\d |{\bm l}_\k|/\d k_\parallel)^2 \gamma_\k^{-1} \rangle_{{\rm FS}i}$,
where ${\bm l}_\k$ is the mean-free-path including the CVC,
and $\theta_\k=\tan^{-1}(l_{\k x}/l_{\k y})$.
Note that $\Delta\rho/\rho_0=-\Delta\s_{xx}/\s_{xx}-(\s_{xy}/\s_{xx})^2$.
Since the CVC enlarge $\d\theta_\k/\d k_\parallel$ near the cold-spot,
both $R_{\rm H}$ and $\Delta\rho/\rho_0$ are enhanced near 
quantum-critical-points.
In addition, $\Delta\rho/\rho_0$ becomes large when ${\bm l}_\k$ is anisotropic
due to the second term in $\Delta\s_{xx}$.
\end{thebibliography}
\end{document}